\documentclass[prd,preprint,showpacs,preprintnumbers,amsmath,amssymb]{revtex4}

 \def\be{\begin{equation}}
 \def\ee{\end{equation}}
 \def\bes{\begin{eqnarray}}
 \def\ees{\end{eqnarray}}
 \def\2{\frac{1}{2}}
 \def\4{\frac{1}{4}}

\catcode`\@=11
\def\@citex[#1]#2{%
\if@filesw \immediate \write \@auxout {\string \citation {#2}}\fi
\@tempcntb\m@ne \let\@h@ld\relax \def\@citea{}%
\@cite{%
  \@for \@citeb:=#2\do {%
    \@ifundefined {b@\@citeb}%
      {\@h@ld\@citea\@tempcntb\m@ne{\bf ?}%
      \@warning {Citation `\@citeb ' on page \thepage \space
undefined}}%
      {\@tempcnta\@tempcntb \advance\@tempcnta\@ne%
      \@tempcntb\number\csname b@\@citeb \endcsname \relax%
      \ifnum\@tempcnta=\@tempcntb %Number follows previous--hold on to
it
        \ifx\@h@ld\relax%
          \edef \@h@ld{\@citea\csname b@\@citeb\endcsname}%
        \else%
          \edef\@h@ld{\ifmmode{-}\else--\fi\csname
b@\@citeb\endcsname}%
        \fi%
      \else%   %  non-successor--dump what's held and do this one
        \@h@ld\@citea\csname b@\@citeb \endcsname%
        \let\@h@ld\relax%
      \fi}%
    \def\@citea{,\penalty\@highpenalty\,}%
  }\@h@ld
}{#1}}

\def\@citeb#1#2{{[#1]\if@tempswa , #2\fi}}
\def\@citeu#1#2{{$^{#1}$\if@tempswa , #2\fi }}
\def\@citep#1#2{{#1\if@tempswa , #2\fi}}

\begin{document}
\preprint{UTHET-08-0401}
                                                                                
\title{Low-lying quasinormal modes of topological AdS black holes and hydrodynamics}

\author{James Alsup}
\email{jalsup1@utk.edu}
\author{George Siopsis}
 \email{siopsis@tennessee.edu}
\affiliation{%
Department of Physics and Astronomy,
The University of Tennessee,
Knoxville, TN 37996 - 1200, USA.
}%
\date{April 2008}%

\begin{abstract}

We analytically calculate the low-lying gravitational quasinormal modes of a topological AdS black hole of arbitrary dimension.
We show that they are in agreement with corresponding results from the hydrodynamics of the gauge theory plasma on the boundary, as required by the AdS/CFT correspondence.
For some of these modes, we obtain a lifetime which is comparable to or longer than the longest lifetime of perturbations of spherical black holes.
Thus, these modes are expected to play an important role in the late time behavior of the gauge theory plasma.
 \end{abstract}

\maketitle

\section{Introduction}

The AdS/CFT correspondence~\cite{adscft,adscftrev} has provided a path for understanding a gauge theory in terms of a dual gravitational description in one higher dimension.  The gauge field in four dimensions is a $\mathcal{N}=4$ super Yang-Mills theory, which is not QCD as there is no confinement among other aspects.  However, once the limit of confinement is breached, $\mathcal{N}=4$ SYM results seem to be relevant to QCD.  Moreover, the plasma seen at RHIC is thought to be strongly interacting \cite{RHICrev} and progress with QCD is difficult to achieve, but AdS/CFT does not rely on the same techniques and is therefore not hampered by the strength of the coupling.  This has cultivated interest in the string theory--RHIC scenarios and progress is rapidly developing \cite{sinzahed,sonstarinets,zahedsin,son}.

Via the AdS/CFT correspondence, information for the plasma in the strong coupling regime is gained by studying the dual theory of a supergravity solution in AdS space.  From the AdS metric one may calculate the stress-energy tensor of the CFT using various techniques \cite{Skenderis}.  The simplest AdS solution to study is the Schwarzschild metric, which has a dual static CFT on the boundary.  This may be extended by looking at small deformations of the Schwarzschild metric, i.e., quasinormal modes which dictate the late-time behavior of the black hole \cite{FGMP}.  Calculating these modes has been held in high importance and thereby studied in vast detail (see \cite{QNM} and references therein).  According to the AdS/CFT correspondence, the lowest frequency modes govern the hydrodynamic behavior of the conformal field theory on the boundary \cite{PSS}.  However, these modes are difficult to find and may be missed by some QNM techniques \cite{Siopsis}.  

In \cite{FGMP,MP} the lowest lying gravitational quasinormal modes for an AdS Schwarzschild solution were numerically calculated in four and five dimensions and were shown to be in agreement with hydrodynamic perturbations of the gauge theory plasma on the AdS boundary.  For AdS$_5$ this was understood as a finite ``conformal soliton flow" after the spherical AdS$_5$ boundary one obtains in global coordinates was conformally mapped to the physically relevant flat Minkowski spacetime.  The perturbations also allowed for calculations of the elliptic flow of the plasma and its thermalization time -- two of the observables at RHIC.  While there is still work to be done, the calculations compared well with what has been found experimentally.

An alternative to a spherical AdS black hole would be to choose one with a hyperbolic horizon \cite{mann,Vanzo:1997gw,Brill:1997mf,Birmingham,Emparan:1999gf}.
They are usually referred to as topological AdS black holes because they possess topologically non-trivial horizons.
Our aim is to elucidate their effect on the gauge theory plasma on the AdS boundary.
By studying gravitational perturbations, we shall show that they possess quasinormal modes whose lifetime is comparable to or longer than their counterparts in the case of horizons with positive curvature (spherical black holes).
These results are in agreement with those obtained by studying the hydrodynamics of the gauge theory plasma on the boundary.
Therefore, the effect of topological AdS black holes must be accounted for in order to understand the behavior of the quark-gluon plasma in heavy ion collisions at RHIC and the LHC via the AdS/CFT correspondence.

In section \ref{sec:2} we discuss the scalar, vector, and tensor gravitational perturbations of a topological AdS black hole in $d$ dimensions.
%Closed, flat, and less studied open boundaries of the Anti-de Sitter space are all analyzed.  
We calculate analytically the lowest lying quasinormal modes using the procedure of ref.~\cite{Siopsis}. In section \ref{sec:3} we study the hydrodynamics of a gauge theory plasma on a hyperbolic space in $d-1$ dimensions extending the results of ref.~\cite{MP}. We show that the frequencies obtained from hydrodynamics are in agreement with their counterparts obtained from black hole perturbations in section \ref{sec:2}. We summarize our conclusions in section \ref{sec:4}.
% These are shown to be in exact agreement with a perturbation of a gauge theory in $d-1$ dimensions.

\section{Topological AdS black holes}
\label{sec:2}

The Einstein equations for vacuum Anti-de Sitter (AdS) space allows for three separate maximally symmetric solutions parameterized with a single parameter $K$
taking the values $0,\pm 1$.
% a choice is made between the three scenarios.  
For $K=0$ we have a flat horizon whereas for
%gives rise to a flat space at the boundary ${\bf R}^{d-2,1}$.  While 
$K=+1$ the horizon is a compact sphere.
The case $K=-1$ yields a horizon which is a hyperbolic space and has been much less studied.
Nevertheless, in the context of the AdS/CFT correspondence all solutions to the Einstein equations should be taken into account. Here we concentrate on the case of black holes with a hyperbolic horizon ($K=-1$) aiming at elucidating their
effect on the gauge theory plasma on the AdS boundary.

The metric of an AdS black hole with $K=-1$ in $d$ spacetime dimensions takes the form
\be\label{metric}
ds^2=- f(r) dt^2+\frac{dr^2}{f(r)}+r^2d\Sigma^2_{d-2}
\ \ , \ \ \ \ f(r) = r^2 -1 - \frac{2\mu}{r^{d-3}} \ee
where we have chosen units in which the AdS radius is $R=1$.
The horizon radius is found from
%and temperature may be found to follow
\be
2\mu=r_H^{d-1}\left( 1-\frac{1}{r_H^2}\right) \ee
The Hawking temperature is
\be\label{eqTH} T_H=\frac{(d-1)r_H^2-(d-3)}{4\pi r_H}
\ee
The area of the horizon is rendered finite by introducing identifications in the hyperbolic space which make the horizon topologically non-trivial.
Thus $\Sigma_{d-2} = \mathbb{H}^{d-2} / \Gamma$ where $\Gamma$ is a dicrete group of isometries of the hyperbolic space $\mathbb{H}^{d-2}$.
Various choices of $\Gamma$ were eagerly studied in the late nineties in preparation for the WMAP as it was thought to possibly describe the type of universe we live in \cite{HyperB}.
E.g., in $d=4$ the boundary may be compactified with periodic boundaries around an octagon specified by $\Gamma$; in higher dimensions the fundamental domain becomes a generalization of the octagon.

The mass and entropy of the hole are given respectively by
%and other black hole thermodynamic quantities may be generalized to the different boundaries with 
\cite{Birmingham}
\be\label{BH}
M=(d-2)(r_H^2-1)\frac{r_H^{d-3}}{16\pi G} V_{d-2}~,~~~ S=\frac{r_H^{d-2}}{4G} V_{d-2}
\ee
where $V_{d-2}$ is the volume of the hyperbolic space $\Sigma_{d-2}$.

For the study of perturbations, we need to understand the behavior of harmonic functions on $\Sigma_{d-2}$. In general, they obey
%By $K$ taking the values $0,\pm 1$ a choice is made between the three scenarios.  The first, $K=0$ gives rise to a flat space at the boundary ${\bf R}^{d-2,1}$.  While $K=1$ corresponds to the spherical boundary  ${\bf S}^{d-2}\times{\bf R}$ as was chosen in \cite{Siopsis,FGMP,MP}.  The last $K=-1$ has a hyperbolic boundary ${\bf H}^{d-2}\times{\bf R}$.  Which has been less studied in the context of AdS/CFT, but nevertheless has the potential to describe a finite plasma when a compactification scheme is chosen \cite{HyperB}.  The compactifications play a large part in determining the quasinormal modes because of the eigenvalues of harmonics in the boundary space.  In general all eigenvalues are positive and the harmonics follow
\be
\left(\nabla^2 + k^2\right){\mathbb T}=0
\ee
Without identifications (i.e., in $\mathbb{H}^{d-2}$), the spectrum is continuous. We obtain \cite{BS}
%where $i$ runs over $d\Sigma_{K,d-2}$
%
%For the spherical case, ${\bf S}^{d-2}$, the eigenvalues are quantized into the typical form
%\be
%k^2=l(l+d-3)-\delta
%\ee
%For the choice of flat and open boundaries the eigenvalue $k^2$ forms a continuous set.  The black hole's hyperbolic boundary allows for the eigenvalues to be parameterized as
\be\label{eqkxi}
k^2=\xi^2+\left(\frac{d-3}{2}\right)^2+\delta
\ee
where $\xi$ is arbitrary and $\delta =0,1,2$ for scalar, vector and tensor perturbations, respectively.
When a compactification scheme is chosen, the spectrum becomes discrete.
Depending on the choice of $\Gamma$, the discretized eigenvalues $\xi$ may be made as small as desired, i.e., zero is an accumulation point of the spectrum of $\xi$ \cite{HyperB}.
As $\xi\to 0$, the complexity of the set of isometries $\Gamma$ increases and the volume $V_{d-2}$ of the hyperbolic space $\Sigma_{d-2}$ diverges (hence also the mass and entropy of the hole).
This ought to be studied numerically for a detailed comparison with experimental data in heavy ion collisions at RHIC through a generalization of the approach of \cite{FGMP}.

Having understood the harmonics on $\Sigma_{d-2}$, we may write the wave equation for gravitational perturbations in the general Schr\"odinger-like form \cite{IK}
\be\label{ME}
-\frac{d^2\phi}{dr_*^2} + V[r(r_*)] \phi = \omega^2 \phi
\ee
in terms of the tortoise coordinate $r_*$ defined by
\be\label{tortoise}
\frac{dr_*}{dr} = \frac{1}{f(r)} \ee
where $f(r)$ is defined in (\ref{metric}).
The potential takes different forms for different types of pertrubation. We shall study each case separately.

\subsection{Vector Perturbations}

The vector potential is given by
\be
V_V=\frac{f(r)}{r^2}\left(k_V^2-1+\frac{(d-2)(d-4)}{4}(r^2-1)-\frac{3(d-2)^2\mu}{r^{d-3}}\right)
\ee
where $k_V^2$ is an eigenvalue of a vector harmonic (eq.~(\ref{eqkxi}) with $\delta = 1$).

It is convenient to introduce the variable
\be u=\left(\frac{r_H}{r}\right)^{d-3} \ee
The wave equation (\ref{ME}) takes the form
\be\label{waveEqn}
-(d-3)^2u^{\frac{d-4}{d-3}}\hat{f}(u)\partial_u\left(u^{\frac{d-4}{d-3}}\hat{f}(u)\partial_u \phi\right)+\hat{V}_V(u)\phi=\hat{\omega}^2\phi
\ee
where
\bes
\hat{V}_V (u)&=&\hat{f}(u)\left[\hat{k}_V^2+\frac{(d-2)(d-4)}{4}u^{\frac{2}{3-d}}-\frac{3(d-2)^2}{4}u\right. \nonumber\\
& & \left. \quad\quad -\frac{1}{r_H^2}\left(1+\frac{(d-2)(d-4)}{4}-\frac{3(d-2)^2}{4}u\right)\right]\nonumber\\
\hat{f}(u)&=& \frac{f(r)}{r^2} =1-u^{\frac{2}{d-3}}\left(u+\frac{1-u}{r_H^2}\right)~,~~~\hat{\omega}^2= \frac{\omega^2}{r_H^2}~,~~~\hat{k}_V^2=\frac{k_V^2}{r_H^2}
\ees
With $\hat{\omega}$ and $\hat{k}_V$ fixed, to leading order in $1/r_H$ this is the same equation as the case of a flat horizon studied in \cite{PSS} and also coincides with the leading order equation in the case of spherical horizon studied in \cite{Siopsis}.
%, and to $\mathcal{O}(1/r_H^2)$ there is only a factor of $K$ difference.  
The curvature of the horizon only comes into play at $\mathcal{O}(1/r_H^2)$, as expected, since the horizon becomes flat in the limit $r_H\to\infty$.
Following the perturbative analysis performed in \cite{Siopsis}, we shall solve the wave equation in the $r_H\to \infty$ limit and add the $\mathcal{O}(1/r_H^2)$ contributions as perturbative corrections (treating $\hat\omega , \hat k_V^2 \sim \mathcal{O} (1/r_H^2)$).

Factoring out the behavior of $\phi$ as it approaches the horizon ($u=1$),
\be
\phi(u)=(1-u)^{-i\frac{\hat{w}}{d-1}}F(u)
\ee
so that the wave equation in the large $r_H$ limit (including $\mathcal{O} (1/r_H^2)$ contributions) becomes
\be\label{sch2} \mathcal{H} F\equiv \mathcal{A} F'' + \mathcal{B} F' + \mathcal{C} F = 0 \ee
where
\bes \mathcal{A} &=& - (d-3)^2 u^{\frac{2d-8}{d-3}} (1-u^{\frac{d-1}{d-3}}) +\frac{2(d-3)^2}{r_H^2} u^2(1-u)\nonumber\\
\mathcal{B} &=& - (d-3) [ d-4-(2d-5)u^{\frac{d-1}{d-3}}]u^{\frac{d-5}{d-3}} - 2(d-3)^2 \frac{i\hat\omega}{d-1}\frac{u^{\frac{2d-8}{d-3}} (1-u^{\frac{d-1}{d-3}})}{1-u} \nonumber\\
& & +\frac{d-3}{r_H^2} u\left[ (d-3)(2-3u) - (d-1) \frac{1-u}{1-u^{\frac{d-1}{d-3}}} u^{\frac{d-1}{d-3}} \right]\nonumber\\
\mathcal{C} &=& \hat{k}_V^2 + \frac{(d-2)[d-4-3(d-2)u^{\frac{d-1}{d-3}}]}{4}u^{-\frac{2}{d-3}} \nonumber\\
& &
- (d-3)\frac{i\hat\omega}{d-1} \frac{[ d-4-(2d-5)u^{\frac{d-1}{d-3}}]u^{\frac{d-5}{d-3}} }{1-u} - (d-3)^2 \frac{i\hat\omega}{d-1}\frac{u^{\frac{2d-8}{d-3}} (1-u^{\frac{d-1}{d-3}})}{(1-u)^2}\nonumber\\
& & - \frac{d-2}{2r_H^2} \left[ d-4-(2d-5)u - (d-1) \frac{1-u}{1-u^{\frac{d-1}{d-3}}} u^{\frac{d-1}{d-3}} \right] \ees
Expanding the wavefunction,
\be F = F_0 + F_1 + \dots \ee
we may solve the wave equation (\ref{sch2}) perturbatively.

The zeroth order wave equation,
\be\label{sch2-0} \mathcal{H}_0 F_0 = 0 \ee
is obtained in the limit $\hat\omega , \hat k_v, 1/r_H^2 \to 0$. Explicitly,
\be\label{sch20} \mathcal{H}_0 F_0= \mathcal{A}_0 F_0'' + \mathcal{B}_0 F_0' + \mathcal{C}_0 F_0 \ee
where
\bes \mathcal{A}_0 &=& - (d-3)^2 u^{\frac{2d-8}{d-3}} (1-u^{\frac{d-1}{d-3}})\nonumber\\
\mathcal{B}_0 &=& - (d-3) [ d-4-(2d-5)u^{\frac{d-1}{d-3}}]u^{\frac{d-5}{d-3}} \nonumber\\
\mathcal{C}_0 &=& \frac{(d-2)[d-4-3(d-2)u^{\frac{d-1}{d-3}}]}{4}u^{-\frac{2}{d-3}}  \ees
The zeroth order wave equation (\ref{sch2-0}) has the two exact solutions
\be
F_0=u^\frac{d-2}{2(d-3)}~,~~~ \check{F}_0=u^{-\frac{d-4}{2(d-3)}}~_2F_1\left(1,-\frac{d-3}{d-1},\frac{2}{d-1};u^{\frac{d-1}{d-3}}\right)
\ee
The former is well behaved at both the horizon ($u\to1$) and the boundary ($u\to0$) but the latter diverges at both ends, therefore it is unacceptable.

The constraint for $\hat{\omega}$ comes from the first order equation which accounts for the $\mathcal{O}(1/r_H^2)$ terms in (\ref{sch2})
\be
\mathcal{H}_0 F_1+\mathcal{H}_1 F_0=0
\ee
solved by
\be
F_1=-F_0\int \frac{\check{F}_0\mathcal{H}_1 F_0}{\mathcal{A}_0\mathcal{W}_0}+\check{F}_0\int \frac{F_0 \mathcal{H}_1 F_0}{\mathcal{A}_0 \mathcal{W}_0}
\ee
where $\mathcal{W}_0$ is the zeroth order wronskian
\be
\mathcal{W}_0 =\frac{1}{u^{\frac{d-4}{d-3}}\left(1-u^{\frac{d-1}{d-3}}\right)}
\ee
The second term in the expression for $F_1$ is ill-behaved at both the boundary and the horizon.
If we choose one of the limits of integration at the boundary ($u=0$),
then the second term becomes regular there.
However, at the horizon it diverges due to the behavior of $\check F_0$. This is avoided if the coefficient of $\check F_0$ vanishes as $u\to 1$.
This requirement yields the constraint
\be\label{eqco} \int_0^1 \frac{F_0\mathcal{H}_1 F_0}{\mathcal{A}_0 \mathcal{W}_0} = 0 \ee
which is a linear equation in $\hat\omega$ whose solution is
\be
\hat{\omega} = -i\frac{\hat{k}_V^2+ \frac{d-3}{r_H^2}}{d-1}
\ee
This is the frequency of the lowest-lying vector quasinormal mode.
It can be written as
\be\label{Vsoln}
\omega 
= -i \frac{\xi^2 + \left(\frac{d-1}{2}\right)^2}{(d-1)r_H} \ee
%  The other solution of the quadratic constraint is the next overtone of the modes and hence unimportant for the discussion of the dual gauge theory's hydrodynamic behavior at the boundary.  Nevertheless, it is an approximation to other studies using the monodromy method for QNMs \cite{QNM}.
This mode is inversely proportional to the radius of the horizon and will dictate the hydrodynamics of the dual gauge theory.
We obtain an upper bound for the lifetime of this mode which may be written in terms of the temperature $T_H \approx \frac{d-1}{4\pi}\, r_H$ (eq.~(\ref{eqTH}) in the large $r_H$ limit and in units in which the AdS radius is $R=1$),
\be \tau = \frac{1}{|\omega|} < \frac{16\pi}{(d-1)^2}\, T_H \ee
In the physically interesting case of $d=5$, this reads $\tau < \pi T_H$. To compare this with the case of a spherical horizon, note that the frequency is given by \cite{Siopsis}
\be\omega^{S^{d-2}} = -i \frac{(l+d-2)(l-1)}{(d-1)r_H} \ee
which yields a maximum lifetime
\be \tau_{\mathrm{max}}^{S^{d-2}} = \frac{4\pi}{d}\, T_H \ee
and in the case $d=5$, we obtain an upper bound of $\frac{4\pi}{5}\, T_H$ which is lower than the upper bound in the hyperbolic case ($\pi T_H$).

\subsection{Scalar Perturbations}

We now turn our attention to scalar perturbations for which the master equation can be cast into the same form as (\ref{waveEqn}) but with a new potential.
\bes\label{ScalarPot}
\hat{V}_S(u)&=&\frac{u^{-\frac{2}{d-3}}-u-\frac{1}{r_H^2}(1-u)}{4(\hat{m}+u)^2}
\Bigg\{ (-6+d) (-4+d) \hat{m}^2-6 (-4+d) (-2+d) \hat{m} u\nonumber\\
&+& (-2+d) d u^2-3 (-6+d) (-2+d) \hat{m}^2 u^{\frac{d-1}{d-3}}\nonumber\\
&+&2 (18+d (-11+2d)) \hat{m} u^{\frac{2(d-2)}{d-3}}+(-2+d)^2 u^{\frac{3d-7}{d-3}}+2 (-2+d) (-1+d) \hat{m}^3 u^{\frac{2}{-3+d}}\nonumber\\
&-&\frac{ u^{\frac{2}{d-3}}}{r_H^2}[(-2+d) \hat{m}^2 (d+2 (-1+d) \hat{m})-3 (-2+d) \hat{m} (-8-6 \hat{m}+d (2+\hat{m})) u\nonumber\\
&+&(24+36 \hat{m}+d
(-10+d-22 \hat{m}+4 d \hat{m})) u^2+(-2+d)^2 u^3]\Bigg\}
\ees
where 
\be\hat{m}=2\frac{k_S^2+d-2}{(d-1)(d-2)(r_H^2-1)}
\ee
and $k_S^2$  is an eigenvalue of a scalar harmonic (eq.~(\ref{eqkxi}) with $\delta = 0$).

A new singularity at $u =-\hat{m}$ arises in the scalar potential.  It is best to factor out the behavior at this point in addition to the behavior at the horizon and boundary.  We see again that the effect of the curvature enters at $\mathcal{O}(1/r_H^2)$ and the wave equation matches the spherical case \cite{Siopsis} at leading order first in $1/r_H$.

Defining 
\be
\phi(u)=(1-u)^{-i\frac{\hat{w}}{d-1}}\frac{u^{\frac{d-4}{2(d-3)}}}{\hat{m}+u}F(u)
\ee
as in the vector case we obtain a wave equation for $F$ which may be solved perturbatively.
In the vector case, we had $\hat\omega, \hat k_V^2 \sim \mathcal{O} (1/r_H^2)$, so keeping terms to $\mathcal{O} (1/r_H^2)$ we could drop terms which were quadratic in $\hat\omega$.
In the scalar case, the frequency has a real part which is related to the speed of sound in the gauge theory fluid \cite{PSS}. In a conformal fluid, the speed of sound is $\frac{1}{\sqrt{d-2}}$. Therefore in the limit $r_H\to \infty$ we expect $\omega\sim \mathcal{O}(1)$, consequently terms which are quadratic in $\hat\omega = \frac{\omega}{r_H}$ must be kept and will contribute at first order in $1/r_H^2$.

The zeroth-order wave equation ought to coincide with the case of a spherical horizon, because the curvature plays no role at leading order. Following \cite{Siopsis}, we choose
\be\label{PerS}
\mathcal{H}_0 F_0=\mathcal{A}_0F_0''+\mathcal{B}_0 F_0' + \mathcal{C}_0 F_0 =0
\ee
where
\bes \mathcal{A}_0 &=& - (d-3)^2 u^{\frac{2d-8}{d-3}} (1-u^{\frac{d-1}{d-3}}) \nonumber\\
\mathcal{B}_0 &=& - (d-3) u^{\frac{2d-8}{d-3}} (1-u^{\frac{d-1}{d-3}}) \left[ \frac{d-4}{u} -\frac{2(d-3)}{\hat m + u} \right]
- (d-3) [ d-4-(2d-5)u^{\frac{d-1}{d-3}}]u^{\frac{d-5}{d-3}} \nonumber\\
\mathcal{C}_0 &=& 0 \ees
%We may treat $\hat{k_S}, \hat{w}$ as small and perturbatively solve the zeroth order wave equation
This zeroth order wave equation has two linearly independent solutions,
\be
F_0=1
\ee
which is well-behaved at all points and a singular one which can be written in terms of the Wronskian,
\be
\check{F}_0=\int \mathcal{W}_0~,~~~
\mathcal{W}_0 =\frac{(\hat{m}+u)^2}{u^{\frac{2d-8}{d-3}}(1-u^{\frac{d-1}{d-3}})}
\ee
Care must be exercised in the case $d=4$ where $\check{F}_0$ does not lead to a singularity at the boundary, however the boundary conditions ought to be altered to Robin boundary conditions \cite{Siopsis,MP}.  

Proceeding as with vector perturbations, a constraint similar to (\ref{eqco}) is found by including terms up to $\mathcal{O} (1/r_H^2)$ which also account for the contributions of $\hat{m} \sim \mathcal{O} (1/r_H^2)$ and $\hat{\omega}\sim \mathcal{O} (1/r_H)$.
After some tedious algebra,
we arrive at a quadratic equation for $\hat{\omega}$,
\be
\frac{d-1}{2}\, \frac{1+(d-2)\hat{m}}{(1+\hat{m})^2}-\frac{1}{r_H^2}\left(\frac{1}{\hat{m}}+\mathcal{O}(1)\right)-i\hat{\omega}\frac{d-3}{(1+\hat{m})^2}-\hat{\omega}^2\left(\frac{1}{\hat{m}}+\mathcal{O}(1)\right)=0
\ee
The two solutions for small $\hat m$ are
\be
\hat{\omega} = \pm\sqrt{\frac{d-1}{2}\hat{m} - \frac{1}{r_H^2}}-i\frac{d-3}{2}\hat{m}
\ee
which may also be written as
\be\label{Ssoln}
\omega_0=\pm\frac{k_S}{\sqrt{d-2}}-i\frac{d-3}{(d-1)(d-2)r_H}\left[k_S^2+d-2\right]
\ee
The real part gives the correct speed of sound ($\frac{1}{\sqrt{d-2}}$) whereas the imaginary part yields the lifetime
\be \tau = \frac{1}{|\Im\omega|} = \frac{4\pi(d-2)}{(d-3)( \xi^2 + \left(\frac{d-1}{2}\right)^2) }\, T_H \ee
This is bounded by
\be \tau < \frac{16\pi(d-2)}{(d-3)( d-1 )^2}\, T_H \ee
to be compared with the maximum lifetime of a scalar mode in the spherical horizon case \cite{Siopsis}
\be \tau_{\mathrm{max}}^{S^{d-2}} = \frac{4(d-2)\pi}{(d-3)d}\, T_H \ee
In the physically interesting case $d=5$, the bound for a hyperbolic horizon is $\frac{3\pi}{2}\, T_H$ which is higher than the maximum lifetime for a spherical horizon, $\frac{6\pi}{5}\, T_H$, as well as the upper bounds of vector modes.

\subsection{Tensor Perturbations}

The remaining quasinormal modes come from tensor perturbations.  Following \cite{IK}, the wave equation may be cast in the same form as (\ref{waveEqn}) with the potential in the large $r_H$ limit,
\be
\hat{V}_T(u)=\frac{d-2}{4}\left(d u^{-\frac{2}{d-3}}-(d-2)u^{\frac{2(d-2)}{d-3}}-2u\right)+\hat{k}_T^2\left(1-u^{\frac{d-1}{d-3}}\right)
\ee
where $\hat{k}_T=k_T/r_H$ and $k_T$ is the tensor harmonic eigenvalue given by eq.~(\ref{eqkxi}) with $\delta = 2$.

The zeroth order wave equation can be solved as in the spherical case \cite{Siopsis} to find the two independent solutions
\be
\phi_0=u^{-\frac{d-2}{2(d-3)}}~,~~~\hat{\phi}_0=u^{-\frac{d-2}{2(d-3)}}\ln\left(1-u^{\frac{d-1}{d-3}}\right)
\ee
Both can be seen to diverge at the boundary ($u\to 0$) and the horizon ($u\to 1$).  Therefore there are no low frequency tensor modes. The lowest modes are expected to have frequencies $\omega\sim \mathcal{O} (r_H)$ and cannot be found using the same perturbative technique as with vector and scalar modes.
We are not interested in finding the tensor modes in this case because they do not contribute to the hydrodynamic behavior of the gauge theory plasma.
%will not work for tensor perturbations.  This is in correspondence with the numerical work in \cite{FGMP} for $K=1,d=5$.

\section{Hydrodynamics}
\label{sec:3}

In the previous section we calculated the lowest lying quasinormal modes whose imaginary part was inversely proportional to the radius of the horizon (and therefore their lifetime was proportional to the Hawking temperature of the black hole).  Based on the analysis in \cite{QNM}, the overtones do not exhibit this behavior; their frequencies are all proportional to the radius of the horizon for large black holes.  This leads to the interpretation of the lowest lying modes corresponding to the hydrodynamics on the dual gauge theory plasma \cite{PSS}, and the subsequent overtones to its microscopic behavior.  In this section, we study the hydrodynamics in the linearized regime of a $d-1$ dimensional fluid with dissipative effects taken into account. The fluid lives on the boundary with topology $\mathbb{R}\times \Sigma_{d-2}$ where $\Sigma_{d-2} = \mathbb{H}^{d-2}/\Gamma$, i.e., the quotient of the hyperbolic space $\mathbb{H}^{d-2}$ with the discrete group of isometries $\Gamma$.
We thus extend earlier results for a spherical boundary \cite{MP}.

Using $\mu,\nu$ running over the boundary with metric
\be ds_{\mathrm{boundary}}^2=-dt^2+d\Sigma_{d-2}^2\ee
and $i,j$ over only the hyperbolic space $\Sigma_{d-2}$, the hydrodynamic equations for the conformal fluid follow in a standard manner,
\bes\label{Stensor}
T^{\mu\nu} &=& (\epsilon+p)u^\mu u^\nu+p g^{\mu\nu}-\eta\left(\triangle^{\mu\lambda}\nabla_\lambda u^\nu+\triangle^{\nu\lambda}\nabla_\lambda u^\mu-\frac{2}{d-2}\triangle^{\mu\nu}\nabla_\lambda u^\lambda\right)-\zeta\triangle^{\mu\nu}\nabla_\lambda u^\lambda\nonumber\\
\nabla_\mu T^{\mu\nu} &=& 0\nonumber\\
T_\mu^\mu &=& 0
\ees
where $\triangle_{\mu\nu}=g_{\mu\nu}+u_\mu u_\nu$ and $\epsilon$, $p$, $\eta$ and $\zeta$ represent the energy density, pressure, shear viscosity and bulk viscosity, respectively, of the conformal field theory.
Two constraints on the parameters immediately follow,
\be\label{eqpz} \epsilon=(d-2)p\ \ , \ \ \ \ \zeta=0\ee
$u^\mu$ is the velocity field of the conformal fluid. The reference frame is chosen so that $u^\mu u_\mu=-1$.  In the rest frame of the fluid, $u^\mu=(1,0,0,0)$. Perturbations introduce small disturbances,
\be
u^\mu=(1,u^i)
\ee
where $u^i$ is small and also allow for small corrections to the pressure so that
\be p=p_0+\delta p\ee
Applying (\ref{Stensor}), we obtain the set of hydrodynamic equations
\bes
0 &=& \nabla_\mu T^{\mu t}=(d-2)\partial_t \delta p+(d-1)p_0 \nabla_i u^i\nonumber\\
0 &=& \nabla_\mu T^{\mu i}=(d-1)p_0 \partial_t u^i +\partial^i\delta p-\eta\left[\nabla^j \nabla_j u^i-(d-3) u^i+\frac{d-4}{d-2}\partial^i (\nabla_j u^j)\right]
\ees
where we used $R_{ij}=-(d-3)g_{ij}$.

Looking first at vector perturbations of the fluid, the appropriate ansatz is \cite{MP}
\be
\delta p=0~,~~~u^i=\mathcal{A}_V e^{-i\Omega t}{\mathbb V}^i
\ee
where ${\mathbb V}^i$ is a vector harmonic.

The first hydrodynamic equation is trivially satisfied and the second becomes
\be\label{Vhydro}
-i \Omega (d-1) p_0+\eta\left[k_V^2+d-3\right]=0
\ee
This can be solved for the frequency $\Omega$ characterizing the deviation from a perfect fluid.  The solution may be written in terms of the parameters of the dual black hole. Using eqs.~(\ref{BH}) and (\ref{eqpz}), we obtain
\be
\frac{\eta}{p_0}=\frac{4\pi\eta}{s}\, \frac{r_H}{r_H^2-1}
\ee
where $s$ is the entropy density. With $\frac{\eta}{s} = \frac{1}{4\pi}$ \cite{PSS} and for large $r_H$ we arrive at the expression for the frequency of vector perturbations
\be
\Omega=-i \frac{k_V^2+d-3}{(d-1) r_H}
\ee
This is in agreement with the frequency of vector modes of the black hole, eq.~(\ref{Vsoln}), on account of the definition (\ref{eqkxi}).

Turning now to scalar hydrodynamic perturbations, we should allow for deviations in pressure as well as the velocity field. The appropriate ansatz is \cite{MP}
\be
u^i=\mathcal{A}_S e^{-i\Omega t}\partial^i \mathbb{S}~,~~~\delta p=\mathcal{B}_S e^{-i\Omega t}\mathbb{S}
\ee
where $\mathbb{S}$ is a scalar harmonic.  The hydrodynamic equations become
\bes
(d-2)i\Omega \mathcal{B}_S+(d-1)p_0 k_S^2\mathcal{A}_S &=& 0\nonumber\\
\mathcal{B}_S+\mathcal{A}_S\left[-i\Omega(d-1)p_0+2(d-3)\eta +2 \eta k_S^2\frac{d-3}{d-2} \right] &=& 0
\ees
This is a linear system of homogeneous equations. To be compatible, their determinant must vanish,
\be \det \left( \begin{array}{cc} (d-2)i\Omega & (d-1) p_0 k_S^2 \\
1 & -i\Omega (d-1) p_0 + 2(d-3)\eta + 2\eta k_S^2 \,\frac{d-3}{d-2} \end{array} \right) = 0 \ee
which imposes a constraint on the frequency $\Omega$.
Working along the same lines as for the vector perturbation, we arrive at the expression for $\Omega$,
\be\label{Shydro}
\Omega=\pm\frac{k_S}{\sqrt{d-2}}-i\frac{d-3}{(d-1)(d-2)r_H}\left[k_S^2+d-2\right]
\ee
which is in exact agreement with the quasinormal frequency of scalar gravitational perturbations (\ref{Ssoln}).

Finally, an ansatz cannot be built to describe tensor perturbations with the associated harmonics because of the tracelessness and zero divergence of tensor spherical harmonics.  This is in consistent with the negative conclusion reached in section \ref{sec:2} on tensor modes of gravitational perturbations of the black hole.

%In addition to comparing our hydrodynamic modes with the quasinormal mode calculation, we may also compare with \cite{MP}.  The hydrodynamics in ${\bf S}^{d-2}\times {\bf R}$ were calculated and our findings (\ref{Vhydro}), (\ref{Shydro}), and the negative tensor result reduce to their findings when $K=1$ is selected.

\section{Conclusion}
\label{sec:4}

We analytically calculated the low-lying quasinormal modes of topological AdS black holes in arbitrary dimension.
These are black holes with hyperbolic horizons of non-trivial topology.
We considered all three different types of perturbations (scalar, vector and tensor) and solved the wave equation \cite{IK} in each case by applying the method of ref.~\cite{Siopsis}.
We obtained quasinormal frequencies which were in agreement with the frequencies obtained by considering perturbations of the gauge theory fluid on the boundary, thus extending results obtained in the case of black holes with spherical horizons \cite{MP}.

In the physically interesting case of five dimensions, we showed that the lifetimes of some of these modes exceed the longest lifetime of the modes of a black hole with spherical horizon \cite{FGMP,Siopsis}.
Therefore, they play an important role in the late time behavior of the gauge theory fluid and ought to be accounted for in order to understand the properties of the quark-gluon plasma produced in heavy ion collisions.
Further work is required for a detailed comparison with experimental data which will determine the topological AdS black hole scenario applicable to RHIC and the LHC.
Following the analysis of \cite{FGMP}, one needs to map the hyperbolic boundary
of the topological black holes onto flat Minkowski space via a conformal map and study the resulting flow of the gauge theory fluid.
Unlike in the case of a spherical boundary, this procedure cannot be carried out analytically for topological AdS black holes owing to the complexity of the (topologically non-trivial) boundary \cite{HyperB}. Instead, one needs to resort to numerical techniques. Work in this direction is in progress.

\section*{Acknowledgment}
Work supported in part by the Department of Energy under grant DE-FG05-91ER40627.
%\newpage

\end{document}